\documentclass[conference]{IEEEtran}
\usepackage{url,hyperref,graphicx}
\usepackage{enumitem}
\usepackage{color}
\usepackage[a4paper]{geometry}
\newgeometry{left=2.471cm,right=2.471cm,top=2.1cm,bottom=2.1cm}
\usepackage{caption,enumitem}
\begin{document}
\hyphenation{threa-te-ned mo-dern mi-cro-bla-ze}
\title{\vspace{-1cm}Towards a hardware-assisted information flow tracking ecosystem for ARM processors}

\author{Muhammad Abdul Wahab$^\alpha$, Pascal Cotret$^\alpha$, Mounir Nasr Allah$^\beta$, Guillaume Hiet$^\beta$\\Vianney Lap\^otre$^\gamma$, Guy Gogniat$^\gamma$\\
$^\alpha$ IETR / SCEE research group, \href{mailto:firstname.lastname@centralesupelec.fr}{firstname.lastname@centralesupelec.fr}\\
$^\beta$ INRIA / CIDRE research group, \href{mailto:firstname.lastname@centralesupelec.fr}{firstname.lastname@centralesupelec.fr}\\
$^\gamma$ Lab-STICC / University of South Brittany, \href{mailto:firstname.lastname@univ-ubs.fr}{firstname.lastname@univ-ubs.fr}\\
}
\maketitle

\begin{abstract}
\noindent
This work details a hardware-assisted approach for information flow tracking implemented on reconfigurable chips. Current solutions are either time-consuming or hardly portable (modifications of both sofware/hardware layers). This work takes benefits from debug components included in ARMv7 processors to retrieve details on instructions committed by the CPU. First results in terms of silicon area and time overheads are also given.
\end{abstract}



\section{Introduction}\vspace{-.1cm}
\noindent Nowadays, high-technology systems are highly threatened by security issues. In the context of software security, original solutions such as DIFT (\emph{Dynamic Information Flow Tracking}) have been proposed since the 2000s. DIFT aims to ensure the application control flow by adding metadata (also known as \emph{tags}) to information containers (e.g. registers, memory addresses). These tags are checked at runtime. DIFT already demonstrated a detection of a wide range of attacks such as SQL injections and buffer overflow.

\noindent However, existing solutions are not widely used in modern SoCs due to hardware and software dependencies. This work provides a clever DIFT implementation for recent SoCs without compromising their security level. This manuscript also describes the internal structure of a new hardware DIFT coprocessor and its implementation results.

\noindent Section \ref{sec:related_work} presents the most relevant related works. Then, Section \ref{sec:objectives} describes the main objectives of this work. Section \ref{sec:current_status} presents the internal mechanisms and implementation results. Finally, Section \ref{sec:conclusion} gives some conclusions and future perspectives.

\section{Related works}\label{sec:related_work}\vspace{-.1cm}
\noindent First and foremost, DIFT implementations were primarily performed in software (without any hardware extensions) as done by Newsome et al. \cite{newsome_05}. However, time overheads were too high (from 300\% up to 3700\%). In order to decrease processing times, several hardware extensions were proposed providing lower penalties at the expense of flexibility (\cite{raksha_07,suh_04,flexitaint_08}).

\noindent Kannan et al. \cite{raksha_09} suggested to separate tags computation from the main application flow: a dedicated coprocessor handles tags, allowing the CPU to run faster. Furthermore, it allows to run simultaneously multiple DIFT checking rules. More recently, other solutions aimed to add features and improve performances shown in \cite{raksha_09}. For instance, Deng et al. \cite{flexcore_10,harmoni_12} proposed a solution to implement DIFT and other similar runtime monitoring techniques such as UMC (\emph{Uninitialized Memory Check}) or BC (\emph{Boundary Check}). 

\noindent Heo et al. \cite{pau_15} proposed a system-level approach to implement DIFT and other related techniques. Information required  by the coprocessor for tags computation is added to the application source code through binary instrumentation. This information is executed at runtime: it sends data from the CPU to a FIFO queue read by the coprocessor. This approach, even though more realistic and generic, presents some drawbacks: 1) information leakage at the interface between the CPU and the coprocessor; 2) code injection attacks may not be detected as the injected code is not instrumented; 3) added instructions through binary instrumentation are architecture-dependent. 

\noindent Table \ref{related_work} is a qualitative comparison of some previous works. \cite{raksha_09,flexcore_10} implemented DIFT using a softcore processor. In both cases, there are modifications of the CPU itself in order to export information. In this work, the main constraint is that the CPU is an ASIC: however, it will be easier to implement on several SoC based on the same architecture. 
\begin{table}[htbp]\vspace{-.25cm}
\centering
\caption{Brief comparison of previous works}\vspace{-.25cm}
\resizebox{\linewidth}{!}{
\begin{tabular}{|l|ccc|}
\hline
\textbf{Approaches} & \textbf{Kannan \cite{raksha_09}} & \textbf{Deng \cite{flexcore_10}} & \textbf{Heo \cite{pau_15}} \\ \hline
Hardcore portability & No & No & Yes\\
Time Overhead & + & ++ & +\\ 
Surface Overhead & + & - & - \\ 
Main CPU & Softcore & Softcore & Softcore \\ \hline
\end{tabular}
}
\label{related_work}
\end{table}

\section{Objectives}
\label{sec:objectives}
\noindent Due to inflexibility and time overheads, DIFT is hardly adopted in modern SoCs. The main goal of this work is to provide a flexible approach for hardware-assisted DIFT based on a standard OS and a heterogeneous architecture such as Xilinx Zynq or Altera DE1-SoC. This work promotes DIFT by proposing a solution with several features:
\begin{itemize}[noitemsep,topsep=0pt]
\item \textbf{Targeting unmodified processors}. Previous works used a softcore LEON3. Zynq devices contain an ARM processor which cannot be modified.
\item \textbf{Scalability}. At first, this work focuses on single-core CPUs. An extension to multicore architectures is planned in the future.
\item \textbf{Efficiency and flexibility}. It must be a low-area and fast solution: the processor must not wait for the coprocessor to complete DIFT tasks (at least, it may halt for the shortest possible time).
\item \textbf{Secure tags computation}. It is assumed that tags and DIFT outputs must not be revealed to an unknown authority.
\end{itemize}

\section{Current status and preliminary results}
\begin{figure}[htbp]
\centering
\includegraphics[width=.75\linewidth]{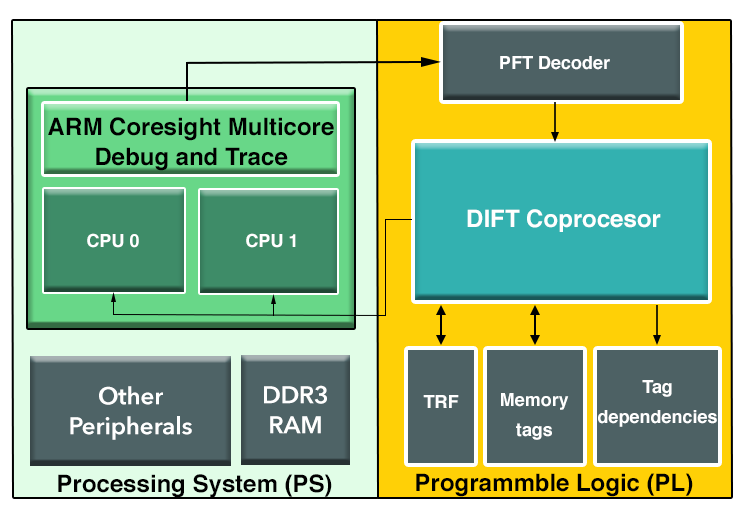}
\caption{Overview of the hardware-assisted DIFT architecture implemented on a Zynq device}
\label{fig:coprocessor}
\end{figure}

\vspace{-.1cm}\label{sec:current_status}%
\noindent The overall architecture used in this work is shown in Figure \ref{fig:coprocessor}. Information required by the coprocessor for tags computation is partially recovered using existing debug components available in ARM processors (also known as Coresight components). Remaining information is obtained through software analysis. 
\vspace*{-.25\baselineskip}
\subsection{Global approach}
\begin{figure}[htbp]
	\centering
	\includegraphics[width=.99\linewidth]{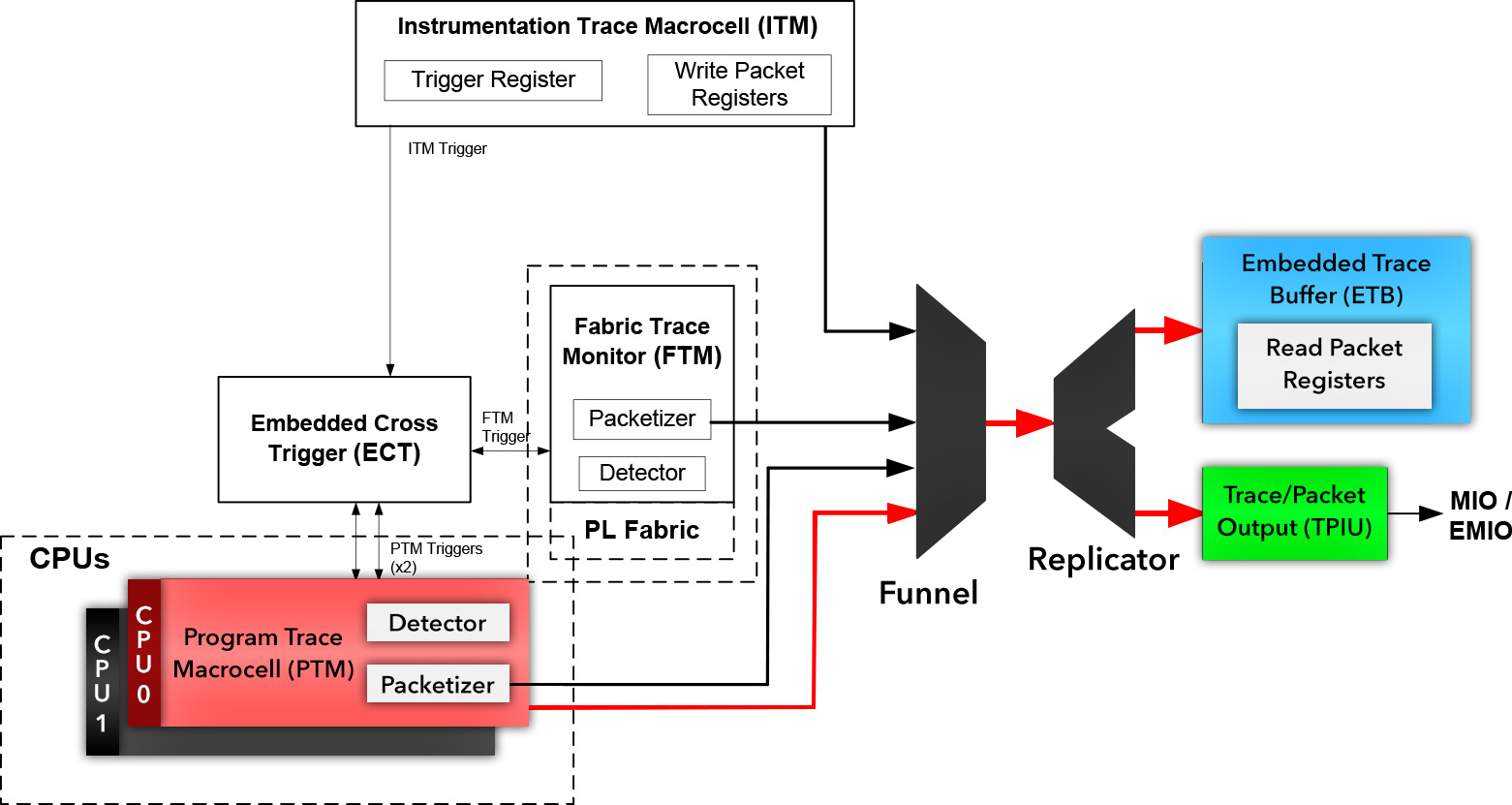}
	\caption{Coresight Components on Zedboard \cite{zynq_trm}}
	\label{fig:Coresight}
\end{figure}\noindent Coresight components (Figure \ref{fig:Coresight}) can be used to debug (or trace) in an efficient manner multicore processors. A PTM (\emph{Program Trace Macrocell}) is assigned to each CPU core: PTMs generate traces (e.g. partial inputs for DIFT computations). Traces only provide runtime information on instructions modifying the program counter (e.g. branches). Traces are transmitted through the funnel and replicator and then pushed in trace sinks (ETB and TPIU). ETB (\emph{Embedded Trace Buffer}) is able to store traces in an 4KB on-chip RAM while TPIU (\emph{Trace Port Interface Unit}) can send it to the programmable logic through the EMIO (Extended Multiplexed I/O) pins.  

\noindent On the PL side, traces are decoded by the PFT decoder (\emph{Program Flow Trace}, see Figure \ref{fig:coprocessor}) and given in a format readable by the DIFT coprocessor. Tag dependencies block contains information obtained through software analysis and rules to handle tags. DIFT coprocessor reads traces given by the PFT decoder and finds which information containers must be propagated. Then, it looks for related tags in TRF (\emph{Tag Register File}) or MR (\emph{Memory tags}): TRF contains tags of each CPU register while MR contains tags for memory locations. The granularity of tags is a user-defined parameter. Finally, the DIFT coprocessor looks for security policy violations and eventually raises an exception.


\subsection{Results}\label{sec:results}
\subsubsection{Traces generation}
\noindent The approach described in this work is at least compatible with SoCs combining an ARM Cortex-A9 processor with a FPGA: Xilinx ZedBoard is the experiment platform in this work. All synthesis were done in Vivado 2014.4. Xilinx Standalone OS was first used to develop Coresight components drivers in order to understand the features offered by such modules and to verify trace contents.

\noindent Coresight drivers for standard Linux are currently being studied and compiled in a Yocto recipe. Traces have been successfully recovered in ETB; however, parasite traces are generated due to context switches. 

\subsubsection{Implementation results}

\noindent For MiBench programs, the overhead introduced by Coresight components is negligible. As tracing components are in hardware and separated from CPU core, almost no overhead is observed. However, the worst case scenario is not evaluated yet and further testing with other benchmarks needs to be done before pronouncing on the efficiency of Coresight components.

\noindent Area results of TRF and PFT Decoder IPs are shown in Table \ref{tab:results}. Percentages are shown relatively to a Microblaze softcore with minimum area configuration (without caches nor BRAMs).

\begin{table}[htbp]
\begin{center}
\caption{IP size for Zedboard (Zynq Z7020)}
\vspace{-.25cm}
\resizebox{\linewidth}{!}{
\begin{tabular}{|l|ccc|}
\hline
\textbf{IP Name} & \textbf{Slice LUTs} & \textbf{Slice Registers} &   \textbf{Slice}\\ \hline
\textbf{\emph{Microblaze}} & 824 & 530 & 300\\ \hline
PFT Decoder & 308 (37\%) & 222 (42\%) & 110(37\%)\\
TRF & 49 (6\%) & 64 (12\%) & 13 (4\%)\\ \hline
\end{tabular}
}
\label{tab:results}
\vspace{-.5cm}
\end{center}
\end{table}

\section{Conclusion and future work}
\label{sec:conclusion}
\noindent A first prototype is currently being developped to demonstrate the feasibility of the approach proposed in this work. Next steps are to build a full-featured system including a secure DIFT coprocessor. Then, DIFT on both Cortex-A9 cores will be implemented by duplicating DIFT coprocessor and other IPs. Dynamic partial reconfiguration will be studied to address energy consumption issues. The proposed approach is not specific to ARM hardcores and may well be adapted to Intel cores using Intel Processor Trace components.
\bibliographystyle{ieeetr}
\bibliography{latex8}
\end{document}